\documentclass[sigconf]{acmart}
\usepackage{booktabs} 
\usepackage{graphicx}
\usepackage{multirow}
\usepackage{amssymb,amsmath,amsfonts,scalerel}
\usepackage{amssymb}
\usepackage{color}
\usepackage{booktabs}
\usepackage{hyperref}
\usepackage[ruled,linesnumbered,vlined]{algorithm2e}
\usepackage{lineno}
\usepackage{subcaption}
\usepackage[textsize=scriptsize,backgroundcolor=green]{todonotes}
\usepackage[inline]{enumitem}

\newlist{inlinelist}{enumerate*}{1}
\setlist*[inlinelist,1]{%
  label=(\roman*),
}

\newcommand{\nosemic}{\renewcommand{\@endalgocfline}{\relax}}
\newcommand{\dosemic}{\renewcommand{\@endalgocfline}{\algocf@endline}}
\let\oldnl\nl
\newcommand{\nonl}{\renewcommand{\nl}{\let\nl\oldnl}}

\newcommand{\partitle}[1]{\vspace{2mm}\noindent\textbf{#1}}

\begin{document}

\title[A Collaborative Ranking Model with Multiple Location-based Similarities for Venue Suggestion]{A Collaborative Ranking Model with \\ Multiple Location-based Similarities for Venue Suggestion}

\author{Mohammad Aliannejadi}
\affiliation{%
  \institution{Universit{\`a} della Svizzera italiana (USI)}
  \city{Lugano} 
  \country{Switzerland}
}
\email{mohammad.alian.nejadi@usi.ch}

\author{Dimitrios Rafailidis}
\affiliation{%
  \institution{University of Mons}
  \city{Mons} 
  \country{Belgium}
}
\email{dimitrios.rafailidis@umons.ac.be}

\author{Fabio Crestani}
\affiliation{%
  \institution{Universit{\`a} della Svizzera italiana (USI)}
  \city{Lugano} 
  \country{Switzerland}
}
\email{fabio.crestani@usi.ch}

\begin{abstract}
Recommending venues plays a critical rule in satisfying users' needs on location-based social networks.
Recent studies have explored the idea of adopting collaborative ranking (CR) for recommendation, combining the idea of learning to rank and collaborative filtering. However, CR suffers from the sparsity problem, mainly because it associates similar users based on exact matching of the venues in their check-in history. Even though research in collaborative filtering has shown that considering auxiliary information such as geographical influence, helps the model to alleviate the sparsity problem, the same direction still needs to be explored in CR. In this work, we present a CR framework that focuses on the top of the ranked list while integrating an arbitrary number of similarity functions between venues as it learns the model's parameters. We further introduce three example similarity measures based on venues' contents and locations. Incorporating cross-venue similarity measures into the model enhances the latent associations between users as similar venues are also taken into account while associating users with each other. Our experiments on the TREC Contextual Suggestion dataset show that our proposed CR model beats other state-of-the-art venue suggestion methods.

\end{abstract}

\copyrightyear{2018} 
\acmYear{2018} 
\setcopyright{acmcopyright}
\acmConference[ICTIR '18]{2018 ACM SIGIR International Conference on the Theory of Information Retrieval}{September 14--17, 2018}{Tianjin, China}
\acmBooktitle{ICTIR '18: 2018 ACM SIGIR International Conference on the Theory of Information Retrieval, September 14--17, 2018, Tianjin, China}
\acmPrice{15.00}
\acmDOI{10.1145/3234944.3234945}
\acmISBN{978-1-4503-5656-5/18/09}

\maketitle

\section{Introduction}
\label{sec-introduction}
With the advent of Location-Based Social Networks (LBSNs), such as Yelp\footnote{\url{https://www.yelp.com/}}, TripAdvisor\footnote{\url{https://www.tripadvisor.com/}}, and Foursquare\footnote{\url{https://www.foursquare.com/}}, users can share check-in data using their mobile devices, together with reviews and other important metadata. 
Being able to help users deal with information overload on LBSNs and provide personalized recommendations is the main purpose of a recommender system~\cite{DBLP:journals/tkde/AdomaviciusT05}.
LBSNs provide a unique opportunity for recommender systems because they collect a myriad of information about venues that is generated by users and was not available before the existence of such services. Such information has been explored through content-based recommendation approaches, and user-generated content such as reviews has also been explored for creating richer user profiles~\cite{alianSigir17}.
 However, many real-world problems limit the accuracy of venue suggestion. For instance, collaborative methods often suffer from the sparsity of users' check-in data. In particular,  even though LBSNs feature a huge number of locations with a large variety, in practice users visit a very limited number of locations, making the user-venue matrix of check-in data extremely sparse~\cite{DBLP:journals/geoinformatica/0003ZWM15}.  
To address the data sparsity problem relevant studies exploit auxiliary information available on LBSNs, such as geographical and temporal information~\cite{DBLP:reference/sp/AdomaviciusT15,DBLP:conf/icadl/BahrainianC17}.

Venue suggestion is often treated as a rating prediction or matrix completion task~\cite{DBLP:conf/nips/SalakhutdinovM07,DBLP:conf/icml/SalakhutdinovM08a} and a large body of research has tried to incorporate contextual similarities into the model while  predicting the ratings. For instance, \citet{DBLP:conf/recsys/KaratzoglouABO10}   proposed an n-dimensional tensor factorization, generalizing matrix factorization to allow for integrating multiple contextual features into the model.
However, \citet{DBLP:conf/wsdm/BalakrishnanC12} pointed out that considering the square loss as a measure of prediction effectiveness is not accurate in the top-$N$ recommendation task. In other words, being able to present a more accurate ranked list to a user when they are only able to check the top-$N$ results should be rewarded. Collaborative Ranking (CR) is based on this idea and focuses on the accuracy of recommendation at the top of the list for each user, by learning the individual's ranking functions in a collaborative manner. However, even tough there has been much work trying to incorporate auxiliary information such as geographical influence in the collaborative filtering framework, incorporating such information in a CR framework is yet to be explored.
 In the literature, it has been shown that the approaches that are based on CR outperform Collaborative-Filtering-based (CF-based) techniques under similar recommendation settings; however, failing to incorporate additional information such as geographical or social influence leads to poorer performance of CR for venue suggestion.

In this work, we propose a novel collaborative venue suggestion framework, called CR-MLS, that enables a model to learn the optimum venue ranking with a focus on the top of the ranked list, while integrating additional information about LBSNs into the model. 
The basic idea behind our proposed method is that the latent association between two users does not necessarily require them to have visited exactly the same venues in the past. On the other hand, if two users have visited very similar venues, we should still be able to use this information to associate those users with each other. In particular, we design the objective function of our CR model to consider the similarity of venues in the loss function with a focus on ranking relevant venues at the top of the recommendation list. After proposing our CR method, we introduce three example cross-venue similarity measures, each of which focuses on a different aspect. We propose a geographical similarity to incorporate the influence of venues in the same neighborhood. Also, we compute a category-based similarity to take into account venues that provide similar services, like serving similar food. We also calculate a review-based similarity score extracting venues' opinion- and context-based similarity. Note that while we introduce three example similarity functions in this paper, our proposed framework essentially is not limited to the number of similarity measures.  The experimental evaluation  shows that considering cross-venue similarities while training the CR model  improves the performance beating all CF and CR state-of-the-art methods. Moreover, as stated in~\cite{Arampatzis:2017:SPV:3146384.3125620}, content-based approaches  generally perform better in cases where the data is extremely sparse. Hence, we also compare the performance of our proposed framework with a state-of-the-art content-based approach.  Observing the difference  between the performance of content-based and collaborative approaches motivated us to explore the combination of our approach with a content-based method. We employed a simple hybrid algorithm outperforming  all state-of-the-art approaches.

This paper's contributions can be summarized as follows:
\begin{itemize}
    \item We introduce a novel CR framework, called CR-MLS, with the focus on the top of the recommendation list, while incorporating the cross-venue similarities into the model.
    \item In order to demonstrate the effectiveness of our CR framework, we propose three different example similarity functions each of which focuses on a different aspect.  
    \item  For the purpose of comparison with the state-of-the-art content-based approaches,  we also propose a simple yet effective hybrid recommendation system, called CR-MLS-Hybrid.
\end{itemize}

The experimental results on data from the TREC Contextual Suggestion track show that our model alleviates the sparsity problem associating similar venues while training the CR model at different settings.


\section{Related Work}
\label{sec-relatedwork}
Here, we briefly review the related work on venue recommendation, collaborative ranking, and contextual suggestion.

\subsection{Venue Recommendation}
Much work has been carried out in venue suggestion based on the core idea that users with similar behavioral history tend to act similarly~\cite{DBLP:journals/cacm/GoldbergNOT92}. This is the underlying idea of CF-based approaches~\cite{koren2008factorization,DBLP:conf/www/SarwarKKR01}. 
CF can be divided into two categories: memory-based and model-based. Memory-based approaches consider user rating as a similarity measure between users or items~\cite{DBLP:conf/www/SarwarKKR01}. Model-based approaches, on the other hand, employ learning strategies like Matrix Factorization (MF)~\cite{koren2008factorization}.
However, CF-based approaches in venue suggestion often suffer from data sparsity since there are a lot of available locations, and a single user can visit only a few of them. As a consequence, the user-venue matrix becomes very sparse, leading to poor performance in cases where there is no significant association between users and venues. 
Many studies have tried to address the data sparsity problem of CF in venue suggestion by exploiting additional information. Ye et al.~\cite{DBLP:conf/sigir/YeYLL11} argued that users check-in behavior is affected by the spatial influence of locations and proposed a unified venue suggestion system incorporating spatial and social influence to address the data sparsity problem. Yin et al.~\cite{DBLP:journals/tois/YinCSHC14} introduced a model that captures user interests as well as local preferences to recommend locations or events to users when they are visiting a new city. Ference et al.~\cite{DBLP:conf/cikm/FerenceYL13} considered user preferences, geographical proximity, and social influences for venue suggestion. Zhang et al.~\cite{DBLP:conf/sigir/ZhangC15} aggregated ratings of users' friends as well as the bias of users on venue categories as power-law distributions. Griesner et al.~\cite{DBLP:conf/recsys/GriesnerAN15} also proposed an approach integrating temporal and geographic influences into MF.  More recently, \citet{DBLP:conf/cikm/ManotumruksaMO17} proposed a deep collaborative  filtering framework with a pairwise ranking function capturing user-venue interactions in a CF manner from sequences of observed feedback by leveraging Multi-Layer Perception and Recurrent Neural Network architectures.
Differently from these studies, our work focuses on ranking more relevant venues on top of the list while learning the users' ranking functions in a collaborative way. Moreover, our model not only considers the geographical influence of neighboring venues but also other contextual similarities.

\subsection{Collaborative Ranking}

Another line of research exploited in this paper lies in combining the ideas of CF and Learning to Rank (LTR).  
LTR methods have been proved to be effective in Information Retrieval (IR)~\cite{DBLP:journals/ftir/Liu09}. LTR learns a ranking function which can predict a relevance score given a query and document. CR takes the idea of predicting preference order of items from LTR and combines it with the idea of learning the loss function in a collaborative way~\cite{DBLP:conf/wsdm/BalakrishnanC12}. \citet{DBLP:conf/nips/WeimerKLS07} used a surrogate convex upper bound of Normalized Discounted Cumulative Gain  (nDCG) error together with MF as the basic rating predictor. \citet{DBLP:conf/cikm/ShiKBLH13} explored optimizing a surrogate lower bound for Expected Reciprocal Rank on data with multiple levels of relevance, while \citet{DBLP:conf/www/Christakopoulou15} followed the idea of pair-wise LTR approaches with an emphasis on the top of the recommendation list. 
\citet{DBLP:conf/cikm/RafailidisC16} introduced two different CR objectives which consider how well the relevant and irrelevant items of users and their social friends have been ranked at the top of the list.
\citet{DBLP:conf/www/LeeBKLS14} assumed that the user-item matrix is low rank within certain neighborhoods of the metric space and minimized a pair-wise loss function. 
\citet{DBLP:conf/recsys/RafailidisC17} proposed a model focusing on trust and distrust between users. They pushed the relevant items of users and their friends at the top of the list, while ranking low those of their foes. \citet{DBLP:conf/www/HuL17} proposed a point-wise CR approach considering user ratings as ordinal rather than viewing them as real values or categorical labels. Also, they emphasized more on positively rated items to improve the performance at the top of recommended list.
None of the works mentioned above consider item-item similarities in the learning phase. On the contrary, we address the data sparsity problem taking into account the similarity of venues in the learning strategy of our model.

\subsection{Contextual Suggestion}
The Contextual Suggestion track (TREC-CS)~\cite{hashemi2016overview}, organized by the Text REtrieval Conference (TREC), aimed to encourage research on context-aware venue suggestion. In this track, the task was to produce a ranked list of venues for each user in a new city, given the user's context and history of preferences in 1-2 other cities. The contextual dimensions were the trip duration, the season, the trip type, and the type of group with whom the user was traveling. These contextual dimensions were introduced in TREC-CS 2015. Since then, among the top runs, few approaches tried to leverage such information. While content-based approaches were consistently among the top runs of the TREC-CS track, \citet{Arampatzis:2017:SPV:3146384.3125620} studied the performance of various content-based, collaborative, and hybrid fusion methods on this task admitting that content-based methods performed best among these methods. \citet{DBLP:conf/trec/Yang015} introduced some handcrafted rules for filtering venues based on their appropriateness to a user's current context.
\citet{alianSigir17,aliannejadi2018personalized} created content-based profiles for users and venues and computed the similarity of a new venue with a user's profile, combined with the score obtained from a venue appropriateness classifier. 
Differently from these works, this study lies on a collaborative approach considering the similarities between venues in the learning process. Inspired from the results of \cite{Arampatzis:2017:SPV:3146384.3125620}, we also combined the results of our approach with a state-of-the-art content-based method.
\section{Proposed Method}\label{sec-method}
Let $\mathcal{P} = \{\rho_1, \dots, \rho_n\}$ and $\mathcal{L} = \{l_1, \dots, l_m\}$ be the sets of $n$ users and $m$ venues, respectively. 
We consider user ratings 1, 2,  and 3 on venues as negative feedback, while ratings 4 and 5 as positive. For each user $\rho_i$, we define $\mathcal{L}^+_i$ as the set of relevant venues, and  $\mathcal{L}^-_i$ as the set of irrelevant ones.
Moreover, let $S_z \in \mathbb{R}^{m \times m}$ be the similarity matrix of venues based on a similarity feature $z$.

We aim at computing a personalized ranking function $f_i(l)$ for each user $\rho_i$ to rank relevant venues higher than irrelevant ones. 
Let $U \in \mathbb{R}^{d \times n}$ be the latent factor of users and $V \in \mathbb{R}^{d \times m}$ be the latent factor of venues, with $\mathbf{u}_i$ and $\mathbf{v}_j$ corresponding to $\rho_i$ and $l_j$, respectively. For user $\rho_i$ the ranking of the venue $l_j$  is computed as follows $f_i(l_j) = \mathbf{u}_i^T\mathbf{v}_j$. The goal of our model is to learn the latent matrices $U$ and $V$.

The rest of the section is organized as follows, first we present the CR model that considers cross-venue similarities to generate venue recommendations, and then we introduce the set of example similarity measures to calculate how close two venues are based on their content and context. Finally, we show how we can combine the ranking of our model with a content-based method resulting in a hybrid model.

\subsection{Collaborative Ranking with Multiple Location-based Similarities}\label{sec-crcs}
In this section, we present our Collaborative Ranking framework, called CR-MLS, to suggest venues for each user $\rho_i$  placing relevant venues at the top of the recommendation list. Our goal is to understand the user's check-in behavior with the  similarities of venues (see Section~\ref{sec-sim}). For example, a user may like all venues that are in the city center and serve pizza. Building ranking functions considering different similarities  between venues also allows us to model latent associations between users with similar tastes who would not be considered in a traditional CR setting. This happens because CR-MLS takes into account the venue similarities as it updates the user and item latent matrices. CR-MLS can build the associations between users as it considers content- and context-based similarities while updating the latent matrices. Notice that our CR-MLS model does not rely on the type of similarity and is not limited to a certain number of similarity features. Hence, it can be a general framework for incorporating any type of similarity features.

We focus on ranking the venues that a user likes higher than the ones she does not. Formally, we aim at ranking venues that belong to $\mathcal{L}^+_i$ higher than those that are in $\mathcal{L}^-_i$. Our goal is to rank the venues with emphasis on the top of the list. 
Let $H_i(l_j^-)$ be the height of an irrelevant venue, that is:

\begin{equation*}
    H_i(l_j^-) = \sum_{k\in\mathcal{L}_i^+} \sum_{z = 1}^{|S|}\Big[\big(\alpha_z \times \mathbf{1}_{[f_i(l_k^+) \leq f_i(l_j^-)]}\big)/S_z(k,j) \Big]~,
\end{equation*}

\noindent
where $\alpha_z$ is the weight of similarity $S_z$ and $\mathbf{1}_{[.]}$ is an indicator function. 
Note that $\alpha_z$ controls the model's bias towards similar venues and can be used to prevent the ``Harry Potter'' problem~\cite{DBLP:conf/ecir/KoolenBBK15}.
Dividing the indicator function by $S_z$ allows the model to incorporate the  similarities into the model while constructing the height for irrelevant items. For example, if an irrelevant item is ranked higher than a relevant item, but they are very similar based on $S_z$, then the denominator will be higher, which means the height of the irrelevant venue will be reduced proportionally. The objective function should aim at minimizing $H_i$ for all irrelevant venues of user $\rho_i$. A lower value of $H_i$ means that there are fewer irrelevant venues ranked higher than relevant ones, and those that are ranked higher are more similar to relevant items. However, indicator functions are not convex and they are not suitable to our optimization strategy. Therefore, we use the logistic loss of the difference between the two functions as a convex upper bound surrogate. We define the difference between the $k^{\text{th}}$ venue and the $j^{\text{th}}$ as follows:
\begin{equation*}
    \delta_i(k,j)=\mathbf{u}_i^T\sum_{z = 1}^{|S|}\big[\alpha_z(\mathbf{v}_k-\mathbf{v}_j)/\exp(|S_z (k,j)|)\big]~.
\end{equation*}

Therefore, the surrogate height function $H_i^{\prime}(l_j^-)$ becomes:
\begin{equation*}
    H_i^{\prime}(l_j^-) = \sum_{k\in \mathcal{L}_i^+}\log\big[1+\exp\big(-\delta_i(k,j)\big)\big]~,
\end{equation*}
\noindent
where $\log(1+\exp(-\delta))$ is the logistic loss of $\delta$. Therefore, the objective function of CR-MLS can be reformulated as a minimization problem with respect to the latent matrices $U$ and $V$ as follows:
\begin{align}
    \label{eq:bla}
    \begin{split}
         & R(U,V)  = \sum_{i=1}^m \frac{1}{n_i}\sum_{j\in\mathcal{L}_i^-}\big(H_i^{\prime}(l_j^-)\big)^2 = \sum_{i=1}^m \frac{1}{n_i} \times \\
         & \sum_{j\in\mathcal{L}_i^-}\Bigg(\sum_{k\in \mathcal{L}_i^+}\log\bigg(1+\exp\Big(-\mathbf{u}_i^T\sum_{z = 1}^{|S|}\big[\alpha_z(\mathbf{v}_k-\mathbf{v}_j)/\exp(|S_z (k,j)|)\big]\Big)\bigg)\Bigg)^2~.
    \end{split}
\end{align}
\noindent For solving the optimization problem of \eqref{eq:bla}, we use a gradient-descent-based alternating optimization algorithm. We first keep $V$ fixed and update $U$, and then keep $U$ fixed and update $V$. Therefore, the update rules of the $t+1$ iteration are:
\begin{equation}
    \label{eq-u-step1}
    \mathbf{u}_i^{t+1} = \mathbf{u}_i^{t} - \gamma \bigtriangledown_{\mathbf{u}_i}R(U^t,V^t), \forall i = 1,\dots,n~,
\end{equation}
\begin{equation}
    \label{eq-v-step1}
    \mathbf{v}_j^{t+1} = \mathbf{v}_j^{t} - \gamma \bigtriangledown_{\mathbf{v}_j}R(U^{t+1},V^t), \forall j = 1,\dots,m~.
\end{equation}
\noindent For reading simplicity we define

\[
\theta(k,j)=\big(1+\exp(\delta(k,j)\big)~.
\]

\noindent The gradients of $R(U,V)$ with respect to $\mathbf{u}_i$ and $\mathbf{v}_j$ are computed as follows:

\begin{align*}
    \begin{split}
    & \bigtriangledown_{\mathbf{u}_i}R(U,V) =  \\
    & = \frac{2}{n_i}\sum_{j\in\mathcal{L}_i^-}\Bigg( H_i^{\prime}(l_j^-) \sum_{k \in \mathcal{L}_i^+} \sum_{z = 1}^{|S|}\bigg[\alpha_z(\mathbf{v}_j - \mathbf{v}_k) / \Big(\exp(|S_z(k,j)|)\theta(k,j)\Big)\bigg]\Bigg)~,
    \end{split}
\end{align*}

\begin{align*}
    \begin{split}
         &  \bigtriangledown_{\mathbf{v}_j}R(U,V) = \\
         & =\sum_{i\in\mathcal{P}^-_j}\frac{2}{n_i}\sum_{h\in\mathcal{L}_i^-}\Bigg( H_i^{\prime}(l_h^-) \sum_{k \in \mathcal{L}_i^+}  \sum_{z = 1}^{|S|} \bigg[ \alpha_z \mathbf{u}_i / \Big(\exp(|S_z(k,h)|)\theta(k,h)\Big) \bigg] \Bigg) \\
        & - \sum_{i\in\mathcal{P}^+_j}\frac{2}{n_i}\sum_{h\in\mathcal{L}_i^-}\Bigg( H_i^{\prime}(l_h^-) \sum_{k \in \mathcal{L}_i^+}  \sum_{z = 1}^{|S|} \bigg[ \alpha_z \mathbf{u}_i / \Big(\exp(|S_z(k,h)|)\theta(k,h)\Big) \bigg] \Bigg)~,
    \end{split}
\end{align*}
\noindent with $\mathcal{P}^+_j$ being the set of users who gave a positive rating to $l_j$ and $\mathcal{P}^-_j$ the set of users who gave a negative rating to $l_j$. Notice that we also consider a regularization term $(\lambda/2)(\|U\|^2 + \|V\|^2)$ to avoid model overfitting in our optimization strategy, where $\lambda$ is the reguralization parameter. Then the final venue recommendations are generated by computing the factorized matrix as the product of $U$ and $V$.

\subsection{Cross-Venue Similarities}
\label{sec-sim}

In this section, in order to demonstrate the effectiveness of our proposed framework, we introduce three example similarity measures. We compute similarity measures between two venues $l_i$ and $l_j$ based on their content and location. Let $S_{ij} = \{S_z(i,j) : z \in \{1,2,3\}\}$ be the set of similarity functions, which are detailed in the following.

\partitle{Geographical similarity.}
First, we compute the geographical similarity between two venues to incorporate the geographical context while characterizing the user's geographical preferences. The similarity is inversely proportional to the distance between two venues.
This score is inspired by the relevant studies~\cite{DBLP:conf/sigir/LiCLPK15,DBLP:conf/kdd/LianZXSCR14} where a simple geographical measure improved the models significantly. 
We use the Haversine formula to compute the angular distance between $l_i$ and $l_j$:
\[
\delta_{ij} = 2 \times \arcsin\Big(\sqrt{\sin^2(\Delta\phi_{ij}/2) + \cos\phi_i \times \cos\phi_j \times \sin^2(\Delta\eta_{ij}/2)}\Big)~,
\]
\noindent where $\phi_i$ and $\phi_j$ are latitudes of $l_i$ and $l_j$ in radian, respectively. Accordingly, $\eta_i$ and $\eta_j$ are longitudes of $l_i$ and $l_j$ in radian. Then we calculate the geographical similarity between $l_i$ and $l_j$ as follows:
\begin{equation}
    S_1(i,j) = \frac{1}{1 + (\delta_{ij} \times R)}~,
    \label{eq-s1}
\end{equation}

\noindent where $R$ is the earth's radius ($R=$6,371KM). 

\partitle{Review-based similarity.}
Online reviews contain a wealth of information about venues as they reflect users' opinions. Since many users explain their context while writing reviews as in, for example: ``I had a quick lunch with my friend right after school,'' it is crucial to measure how similar two venues are in terms of the reviews they received. It is also important to consider how a particular user rated venues that are similarly reviewed by others. Therefore, we train a Support Vector Machine (SVM) classifier with linear kernel to estimate the review-based similarity. 
The choice of SVM classifier was highly inspired by observing its notable performance in other studies~\cite{DBLP:conf/sac/AliannejadiMC17,DBLP:journals/corr/abs-1803-08354}.
For each venue, we train a different SVM classifier. We take the positive reviews of the corresponding venue as positive training samples and the negative reviews as negative training samples. We denote the trained SVM classifier of $l_i$ as SVM$_i$. Notice that the reviews used for training are independent of a particular user's reviews about a specific venue. In other words, we train the SVM classifiers using the online public reviews available on LBSNs. Finally, we compute the review-based similarity between $l_i$ and $l_j$ by classifying the reviews of $l_j$ using SVM$_i$. Note that we use both positive and negative reviews of $l_j$ to classify $l_j$ with SVM$_i$. The value of SVM$_i$'s decision function computes the similarity of two venues $l_i$ and $l_j$, denoted as $S_{2}(i,j)$.

\partitle{Category-based similarity.}
While it is essential to exploit users' ratings considering geographical proximity and review-based similarity, it is also crucial to take into account how users rate venues that are similar in terms of their categories. For example, a user who likes pizza is more likely to visit a pizza place and rate it positively. It has been shown in relevant works that incorporating venue categories into the recommender system is crucial \cite{DBLP:conf/sigir/ZhangC15}.
We calculate the cosine similarity between the vectors of categories associated with venues $l_i$ and $l_j$ on LBSNs as follows:
\begin{equation}
    S_{3}(i,j) = \frac{\mathbf{c}_i.\mathbf{c}_j}{\|\mathbf{c}_i\|_2 \|\mathbf{c}_j\|_2}~,
    \label{eq-s3}
\end{equation}

\noindent where $\mathbf{c}_i$ and $\mathbf{c}_j$ are the category vectors for $l_i$ and $l_j$, respectively.

\subsection{System Overview}
\label{sec-overview}
Algorithm~\ref{algo-crcs} summarizes our proposed CR-MLS with the three cross-venue similarity measures that we introduced in Section~\ref{sec-sim}. As we can see from line 1 to 5, the three similarity scores are computed for all the pairs of venues and stored in a three-dimensional matrix called $S$. Then, from line 7 to 13, the main steps of CR-MLS are done to learn the parameters of the model, taking the similarity of venues into account.

\partitle{Efficiency.} Note that one could argue that computing the similarity measures between all pairs of venues is not efficient. Although this is a valid argument, it is worth noting that the main focus of our paper is not on efficiency. However, we believe that a more efficient strategy for selecting venue pairs could be studied to improve the complexity of computing $S$. 

\begin{algorithm}[pt]
\DontPrintSemicolon
\KwIn{ $\mathcal{P}$, $\mathcal{L}$, $maxIter$, $\mathbf{c}$, $\phi$, $\eta$, $\{d, \lambda, \alpha, \epsilon\}$} 
\KwOut{$U$, $V$} 

 \ForAll{$i \in |\mathcal{L}|$}{
    \ForAll{$j \in |\mathcal{L}|$}{
        \quad $S_1(i,j) \gets 1/(1+(\delta_{ij}\times R))$~(Equation \eqref{eq-s1}) \\
        \quad $S_2(i,j) \gets$ value of decision function of SVM$_i$ given reviews of $l_j$ \\
        \quad $S_3(i,j) \gets (\mathbf{c}_i.\mathbf{c}_j)/(\|\mathbf{c}_i\|_2\|\mathbf{c_j}\|_2)$~(Equation \eqref{eq-s3}) \\
    }
}

$t \gets 0$ \\
Initialize $U^{t+1}$, $V^{t+1}$ \\
$\theta^{t+1} \gets R(U^{t+1},V^{t+1})$, $\theta^t = \theta^{t+1}/2$ \\

\While{($abs(\theta^{t+1} - \theta^{t}) > \epsilon$) $\land$ ($t < maxIter$)} {
    \quad $t \gets t+1$ \\
	\quad Update $\mathbf{u}_i^{t+1}, \forall i = 1,\dots,n$~(Equation \eqref{eq-u-step1}) \\
	\quad Update $\mathbf{v}_j^{t+1}, \forall j = 1,\dots,m$~(Equation \eqref{eq-v-step1}) \\	
	\quad $\theta^{t+1} \gets R(U^{t+1},V^{t+1})$ \\
\nonl\bf{end}
}
$U \gets U^{t+1}, V \gets V^{t+1}$
\caption{{Collaborative Ranking with Multiple Location-based Similarities Algorithm (CR-MLS)}}
\label{algo-crcs}
\end{algorithm}

\subsection{Hybrid Venue Suggestion}
\label{sec-hybrid}
In this section, we combine the output of CR-MLS with the output of a state-of-the-art content-based method called LinearRankRev~\cite{alian2015}. Our goal is to demonstrate the effectiveness of our approach when combined with a content-based approach on a highly sparse dataset. To this aim, we first produce the ranking using both methods and consider the ordinal position of a venue as its score. For example, the first venue in a ranked list gets the score of 1 and the score of the second one becomes 2. Let $Rk_{S}(\rho_i,l_j)$ be the ordinal position of venue $l_j$ for user $\rho_i$ using CR-MLS and $Rk_{L}(\rho_i,l_j)$ be the ordinal position of $l_j$ for $\rho_i$ using LinearRankRev. We calculate the linear combination of the two ranked lists as follows:
\[
Rk(\rho_i,l_j) = \beta \times Rk_{S}(\rho_i,l_j) + (1-\beta) \times Rk_{L}(\rho_i,l_j)~,
\]
where $\beta$ is the combination weight. The final ranking is obtained by sorting the venues in terms of $Rk$. In the following section, we call the results of this model CR-MLS-Hybrid.

\begin{table*}[t]
\centering
\caption{Performance evaluation on TREC-CS in terms of P@k and nDCG@k with k $\in \{1,2,3,4,5\}$. Bold values denote the best scores compared with collaborative approaches and the content-based approach separately. The superscript $\dagger$ denotes significant improvements compared to all collaborative baselines and $\ddagger$ denotes significant improvements compared to the content-based baseline (i.e., LinearRankRev), for $p<$0.05 in paired t-test.}
\begin{tabular}{l@{\quad}l@{\quad}l@{\quad}l@{\quad}lll@{\quad}l@{\quad}l@{\quad}l@{\quad}l@{\quad}l}
\toprule
 & P@1 & P@2 & P@3 & P@4 & P@5  && nDCG@1 & nDCG@2 & nDCG@3 & nDCG@4 & nDCG@5 \\
\cmidrule{2-6} \cmidrule{8-12}
P-Push & 0.5635 & 0.5179 & 0.5079 & 0.4772 & 0.4524 && 0.5635 & 0.5282 & 0.5188 & 0.4963 & 0.4775 \\
RH-Push & 0.4606 & 0.4547 & 0.4580 & 0.4626 & 0.4567 && 0.4606 & 0.4561 & 0.4581 & 0.4611 & 0.4575 \\
IRenMF & 0.5037 & 0.4706 & 0.4767  & 0.4706 & 0.4610 && 0.5037 & 0.4781 & 0.4806 & 0.4759 & 0.4689 \\ 
GeoMF & 0.4743 & 0.4871 & 0.4714  & 0.4789 & 0.4740  && 0.4743 & 0.4842 & 0.4879  & 0.4801 & 0.4774  \\
Rank-GeoFM & 0.5662 & 0.5441 & 0.5392  & 0.4926 & 0.4743 && 0.5662 & 0.5491 & 0.5445 & 0.5123 & 0.4976 \\
CR-MLS & \textbf{0.6605}$^{\dagger\ddagger}$ & \textbf{0.5830}$^\dagger$ & \textbf{0.5510}$^\dagger$ & \textbf{0.5055} & \textbf{0.4804} && \textbf{0.6605}$^{\dagger\ddagger}$ & \textbf{0.6043}$^\dagger$ & \textbf{0.586}5$^\dagger$ & \textbf{0.5614}$^\dagger$ & \textbf{0.5509}$^\dagger$ \\

\midrule
\midrule
LinearCatRev & 0.6471 & 0.6452 & 0.6336 & 0.6121 & 0.5868 && 0.6471 & 0.6498 & 0.6499 & 0.6444 & 0.6394 \\
CR-MLS-Hybrid & \textbf{0.6801}$^{\dagger\ddagger}$ & \textbf{0.6673}$^{\dagger\ddagger}$ & \textbf{0.6458}$^\dagger$ & \textbf{0.6140}$^\dagger$ & \textbf{0.5919}$^\dagger$ && \textbf{0.6801}$^{\dagger\ddagger}$ & \textbf{0.6734}$^{\dagger\ddagger}$ & \textbf{0.6672}$^{\dagger\ddagger}$ & \textbf{0.6562}$^\dagger$ & \textbf{0.6538}$^{\dagger\ddagger}$ \\
\bottomrule
\end{tabular}
\label{tab-pk}

\end{table*}

\section{Experimental Evaluation}\label{sec-experiments}
In this section, we first introduce the experimental setup describing the dataset, evaluation metrics, parameter tuning as well as compared methods. Then we present the results together with detailed discussions.

\subsection{Setup}
\partitle{Dataset.} 
We evaluate our approach on a benchmark dataset, made available by the TREC. The dataset is the combination of the data for the TREC-CS 2015 and 2016 tracks~\cite{hashemi2016overview}. Since TREC released the ground truth for 211 and 58 users in TREC-CS 2015 and 2016, respectively; and the settings for both datasets were identical, we combined both datasets to generate a single larger dataset, denoted as \textit{TREC-CS}. In doing so, the sparsity of the user-venue matrix is increased.
The task was to produce a ranked list of venues in a new city for users given their history of venue preferences in other cities. Each user has visited and rated 30 to 60 venues in one or two cities. We used the publicly available crawls of~\cite{DBLP:conf/sigir/AliannejadiMC17} as additional information. More specifically, we used additional information from Yelp such as reviews, categories, and address. We then used HERE API\footnote{\url{https://developer.here.com/}} to extract geographical coordinates given a venue's address. In summary, the unified TREC-CS dataset consists of 269 users. The auxiliary information was crawled from Yelp for 6,346 venues. The average number of reviews per venue is 105.55 and the average number of categories per venue is 2.44.

\partitle{Evaluation protocol.}
We use the official evaluation metrics of TREC for this task, that is, P@k (Precision at k) and nDCG@k with k $\in \{1,2,3,4,5\}$.
Since our approach exploits the influence of neighboring venues, evaluating recommendation in a new city where the user does not have any check-in records does not allow us to study the effect of the geographical similarity function. Hence, we evaluate our method in the same way as the state-of-the-art approaches evaluated their works~\cite{DBLP:conf/sigir/LiCLPK15}, that is, we use 70\% of the check-in data as \textit{training set} (17.9K ratings), 10\% as \textit{validation set} (2.5K ratings), and 20\% as \textit{testing set} (5.4K ratings). Notice that since the ratings are not timestamped, we split the dataset randomly; hence we repeat our experiments 5 times and report the average P@k and nDCG@k.

\partitle{Parameter tuning.}
In this paper, we set $\gamma= 0.0001$ for the learning rate to ensure the generalization of our model, and then we tune the other parameters based on the validation set. We find the optimal values for the parameters using the validation set and use them in the test set. The best performance of CR-MLS-Hybrid is achieved with $\beta=0.2$. We discuss the effect of the other parameters in Section~\ref{sec-res}.

\partitle{Compared methods.}
We compare our CR-MLS and CR-MLS-Hybrid models with approaches that consider ranking and geographical influence for venue suggestion and approaches based on collaborative ranking with emphasis on the ranking performance at the top of the list. 
We also compare our models with the TREC's best performing run.
Thus, we compare CR-MLS and CR-MLS-Hybrid with the following methods:\\
$-$ Collaborative methods:
\begin{itemize}
    \item \textbf{P-Push}~\cite{DBLP:conf/www/Christakopoulou15} focuses on the ranking performance at the top of the list using a $p$-norm height function in CR. P-Push does not include any contextual information in its learning strategy.
    \item \textbf{RH-Push}~\cite{DBLP:conf/www/Christakopoulou15} is another push CR model based on reverse height, focusing on the ranking performance at the top of the list. RH-Push does not consider any contextual information in its learning strategy either.
    \item \textbf{IRenMF}~\cite{DBLP:conf/cikm/LiuWSM14} is based on weighted MF~\cite{DBLP:conf/icdm/PanZCLLSY08} exploiting two
levels of geographical neighborhood characteristics: nearest neighboring locations share more similar user preferences, while locations in the same geographical region may share similar user preferences.
    \item \textbf{GeoMF}~\cite{DBLP:conf/kdd/LianZXSCR14} augments users' and venues' latent factors in the factorization model with activity area vectors of
users and influence area vectors of venues, respectively.
    \item \textbf{Rank-GeoFM}~\cite{DBLP:conf/sigir/LiCLPK15} is the state-of-the-art venue recommendation algorithm. It is a ranking-based MF model that includes the geographical influence of neighboring venues while learning users' preference rankings for venues.
\end{itemize}
$-$ Content-based method:
\begin{itemize}
    \item \textbf{LinearCatRev}~\cite{alian2015,alianTREC2016} is the best performing model of TREC-CS 2015 and 2016. It is a content-based recommender system which extracts information from different LBSNs and uses it to calculate category-based and review-based scores. Then, it combines the scores using linear interpolation. The main difference between LinearCatRev and other methods is that it is a content-based method focusing on creating rich user and venue profiles. Although this method performs very well on this dataset, there are major concerns regarding its scalability, mainly because it trains a separate classifier per user, something that can be challenging in a real-life recommendation scenario.
\end{itemize}
We compare the performance of our proposed models with these methods in the following section.
\subsection{Results and Discussion}
\label{sec-res}

\begin{table*}[t]
\centering
\caption{Effect on P@5 and nDCG@5 of different number of venues that users visited as training set.. The superscript $\dagger$ denotes significant improvements compared to all collaborative baselines and $\ddagger$ denotes significant improvements compared to the content-based baseline (i.e., LinearRankRev), for $p<$0.05 in paired t-test.}
\begin{tabular}{l@{\quad}l@{\quad}l@{\quad}ll@{\quad}l@{\quad}l@{\quad}l}
\toprule
 & \multicolumn{3}{c}{P@5} && \multicolumn{3}{c}{nDCG@5} \\
\cmidrule{2-4} \cmidrule{6-8}
 Number of venues & \multicolumn{1}{l}{40} &  \multicolumn{1}{l}{50} & \multicolumn{1}{l}{60} && \multicolumn{1}{l}{40} & \multicolumn{1}{l}{50} & \multicolumn{1}{l}{60}  \\
\cmidrule{2-4} \cmidrule{6-8}
P-Push & 0.4278 & 0.4346 & 0.4466 && 0.4493 & 0.4375 & 0.4744 \\
RH-Push & 0.4343 & 0.4556 & 0.4574 && 0.4277 & 0.4659 & 0.4606 \\
IRenMF & 0.4404 & 0.4588 & 0.4588 && 0.4485 & 0.4654 & 0.4636 \\ 
GeoMF & 0.4544 & 0.4618 & 0.4640 && 0.4559 & 0.4573 & 0.4636  \\
Rank-GeoFM & 0.4551 & 0.4727 & 0.4728 && 0.4576 & 0.5006 & 0.5027 \\
CR-MLS & \textbf{0.4677}$^{\dagger}$ & \textbf{0.4732} & \textbf{0.4800} && \textbf{0.4736}$^{\dagger}$ & \textbf{0.5302}$^{\dagger}$ & \textbf{0.5450}$^{\dagger}$ \\
\midrule
\midrule
LinearCatRev & 0.5706 & 0.5632 & 0.5721 && 0.6246 & 0.6195 & 0.6260 \\
CR-MLS-Hybrid & \textbf{0.5772}$^{\dagger}$ & \textbf{0.5787}$^{\dagger\ddagger}$ & \textbf{0.5853}$^{\dagger\ddagger}$ && \textbf{0.6357}$^\dagger$ & \textbf{0.6353}$^{\dagger\ddagger}$ & \textbf{0.6433}$^{\dagger\ddagger}$ \\
\bottomrule
\end{tabular}
\label{tab-nvenues}

\end{table*}

\partitle{Comparison with collaborative state-of-the-art.}
\autoref{tab-pk} reports the performance of all the models on TREC-CS in terms of P@k and nDCG@k with k $\in \{1,2,3,4,5\}$. We observe that the push CR-based models perform worse in terms of P@5 and nDCG@5. This occurs because the baseline push CR-models do not consider any similarities while training the model. Although P-Push performs more effectively than RH-Push regarding P@1-4, it has the worst performance in terms of P@5 among all models because P-Push focuses on optimizing the model for the negative items, while RH-Push focuses on the positive items. However, since none of them take into account the similarities between venues, they cannot perform as well as the other models. 
Among the methods that consider geographical influence in the model, Rank-GeoFM performs better. This is because Rank-GeoFM considers venue suggestion problem as a ranking problem, similar to our approach. Rank-GeoFM, IRenMF, and GeoMF perform better than CR-based baselines indicating that geographical influence is an important factor in venue suggestion. While Rank-GeoFM and GeoMF perform similarly in terms of P@5, we observe that Rank-GeoFM performs better in term of nDCG@5 indicating that a ranking-based approach enables a system to rank more relevant items higher in the ranking. 

Our proposed CR-MLS model significantly outperforms all collaborative state-of-the-art methods in terms of P@1-3 and nDCG@1-5 (according to pairwise t-test at $p < 0.05$). Compared to the state-of-the-art method, Rank-GeoFM, the improvements in terms nDCG@1 and nDCG@5 are 17\% and 11\%, respectively. 
This indicates that our proposed CR-MLS can address the data sparsity problem by incorporating different types of similarities. While the geographical similarity includes the neighborhood influences in the model, the category-based similarity takes into account users with similar tastes when they do not share the same check-in records. In addition to that, the review-based similarity models venues similarities in terms of other users' opinions in various contexts. Fusing these similarity measures with a CR-based model enables CR-MLS to form complicated similarity affinities among venues and propagate it to the users. Hence, our proposed CR-MLS addresses the data sparsity problem more effectively than other state-of-the-art models, indicated by the high recommendation accuracy. Finally, more improvements in terms of nDCG@k suggests that CR-MLS is able to rank higher the venues that are rated higher by the users.

\partitle{Comparison with content-base state-of-the-art.} We also compare the performance of our model when combined with a content-based model as a hybrid method, called CR-MLS-Hybrid (see Section~\ref{sec-hybrid}). We see in \autoref{tab-pk} that LinearCatRev performs better than all collaborative approaches. This result is inline with the findings of \citet{Arampatzis:2017:SPV:3146384.3125620}, that is, due to high sparsity of TREC-CS dataset content-based approaches are generally more effective than collaborative methods. However, we observe that CR-MLS exhibits a better performance in terms of nDCG@1 compared to LinearCatRev. This motivated us to combine the ranking of CR-MLS and LinearCatRev to build a stronger hybrid approach. As we can see in \autoref{tab-pk}, CR-MLS-Hybrid outperforms all other methods. In particular, we observe more improvements in terms of nDCG@k, indicating that CR-MLS-Hybrid is also able to rank higher the venues with higher rating. It is also worth noting that significant improvement from LinearCatRev indicates that CR-MLS not only performs better than state-of-the-art collaborative methods, but also is able to exploit user-venue associations in a way that the content-based approach fails to do.

\partitle{Impact of number of visited venues.}
\autoref{tab-nvenues} shows P@5 and nDCG@5 of all models when varying the number of venues that each user has visited in the training set. \autoref{tab-nvenues} shows that CR-MLS achieves the highest accuracy, compared to the other collaborative models and CR-MLS-Hybrid compared to all other models for all different number of venues. 
This result indicates that CR-MLS can address the sparsity problem better when the training set is smaller. Also, we observe a more robust behavior of CR-MLS compared to the baselines suggesting that incorporating similarities enables the model to deal with noise and data sparsity more effectively. This is more obvious when observing that CR-MLS outperforms all the baseline methods with a larger margin in terms of nDCG@5.
Also, we can see that LinearCatRev's performance is less robust as we vary the number of venues. We do not observe the same behavior in CR-MLS-Hybrid's performance implying that combining the ranking of CR-MLS with LinearCatRev also improves the stability of the content-based approach when trained with less venues in the training set.

\begin{figure*}[t]
    \centering
    \begin{subfigure}[b]{0.25\textwidth}
        \includegraphics[width=\textwidth]{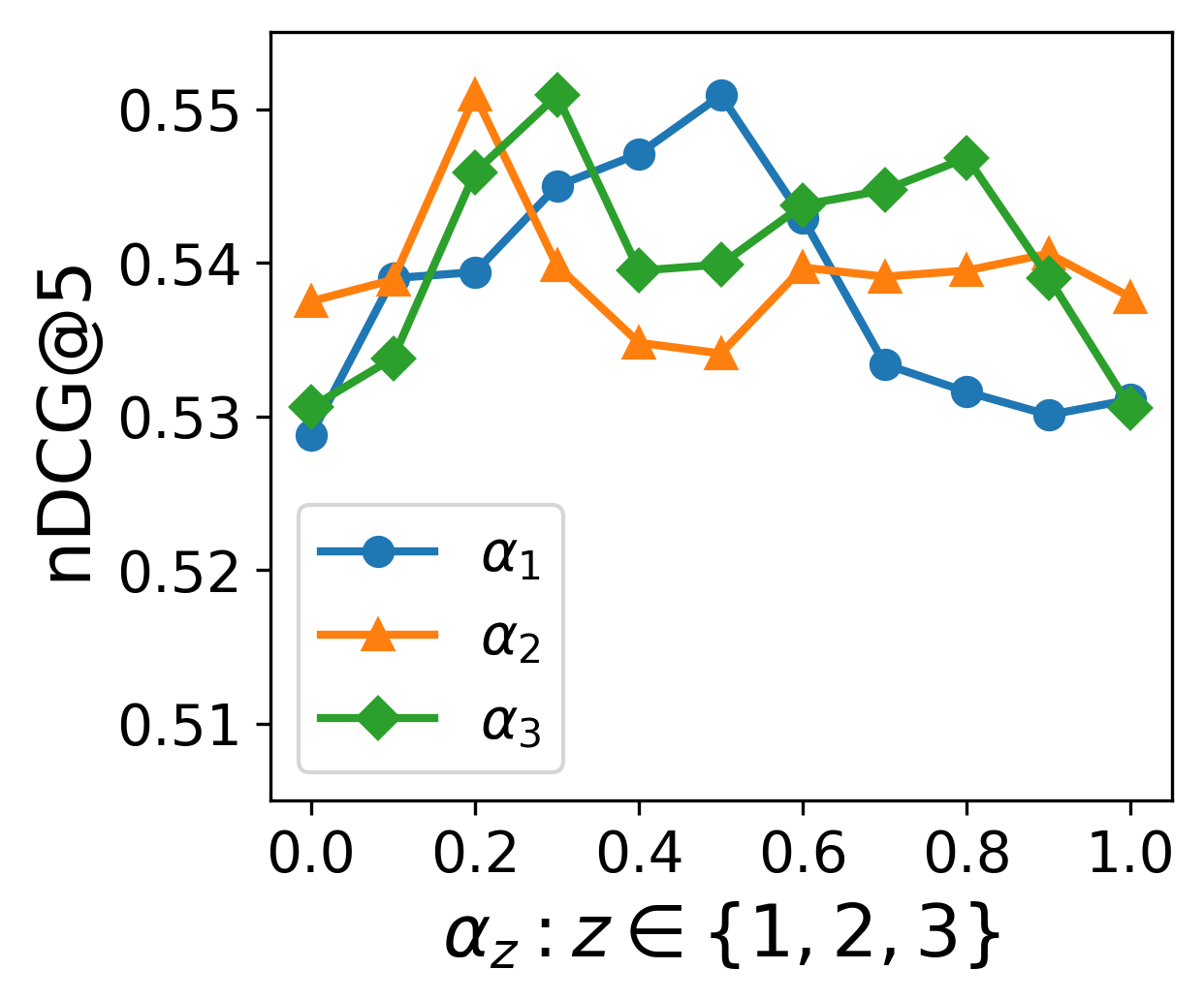}
        \caption{Effect of $\alpha_1$, $\alpha_2$, $\alpha_3$}
        \label{fig-weights}
    \end{subfigure}
    ~ 
    \begin{subfigure}[b]{0.25\textwidth}
        \includegraphics[width=\textwidth]{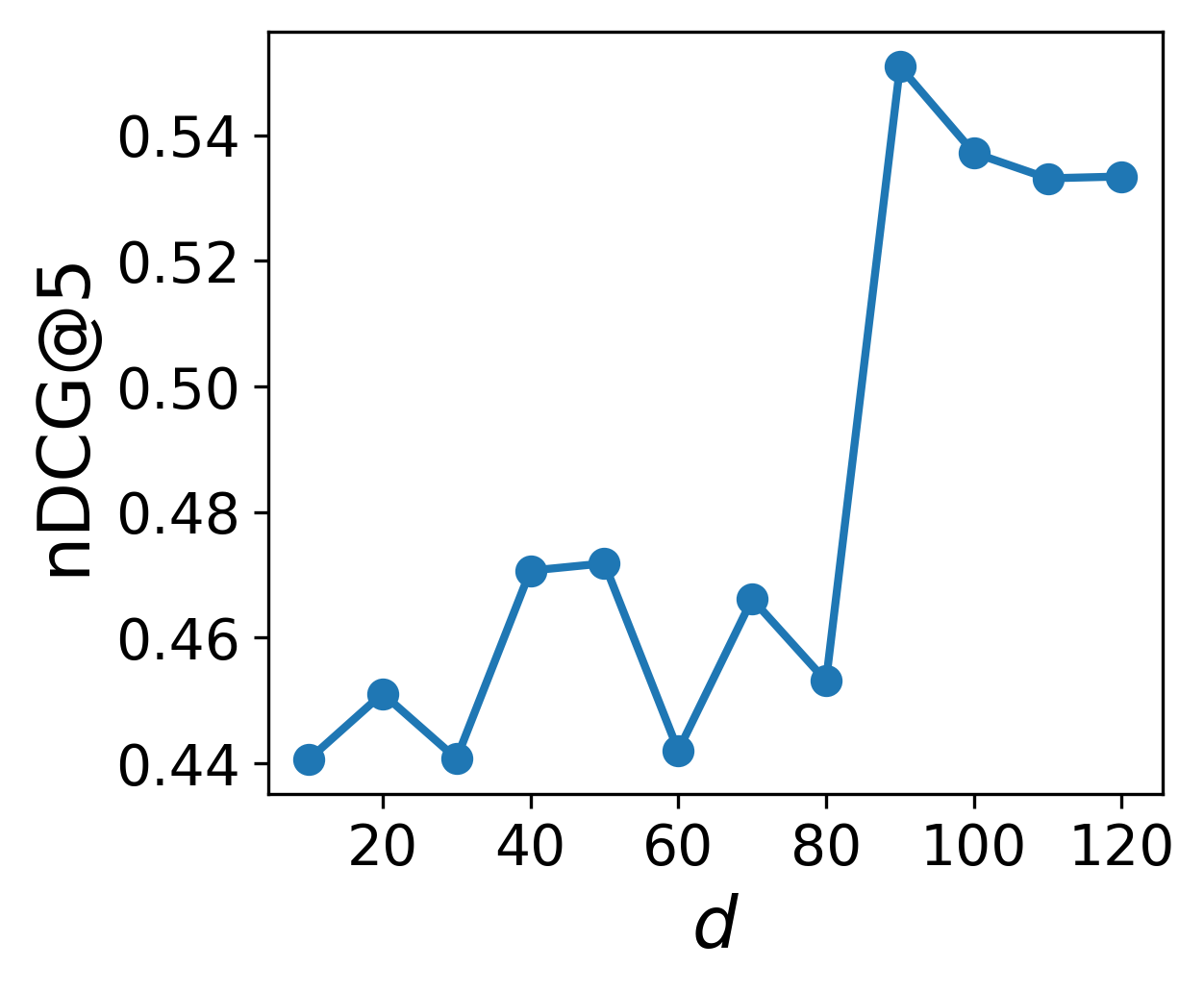}
        \caption{Effect of $d$}
        \label{fig-d}
    \end{subfigure}
    ~ 
    \begin{subfigure}[b]{0.25\textwidth}
        \includegraphics[width=\textwidth]{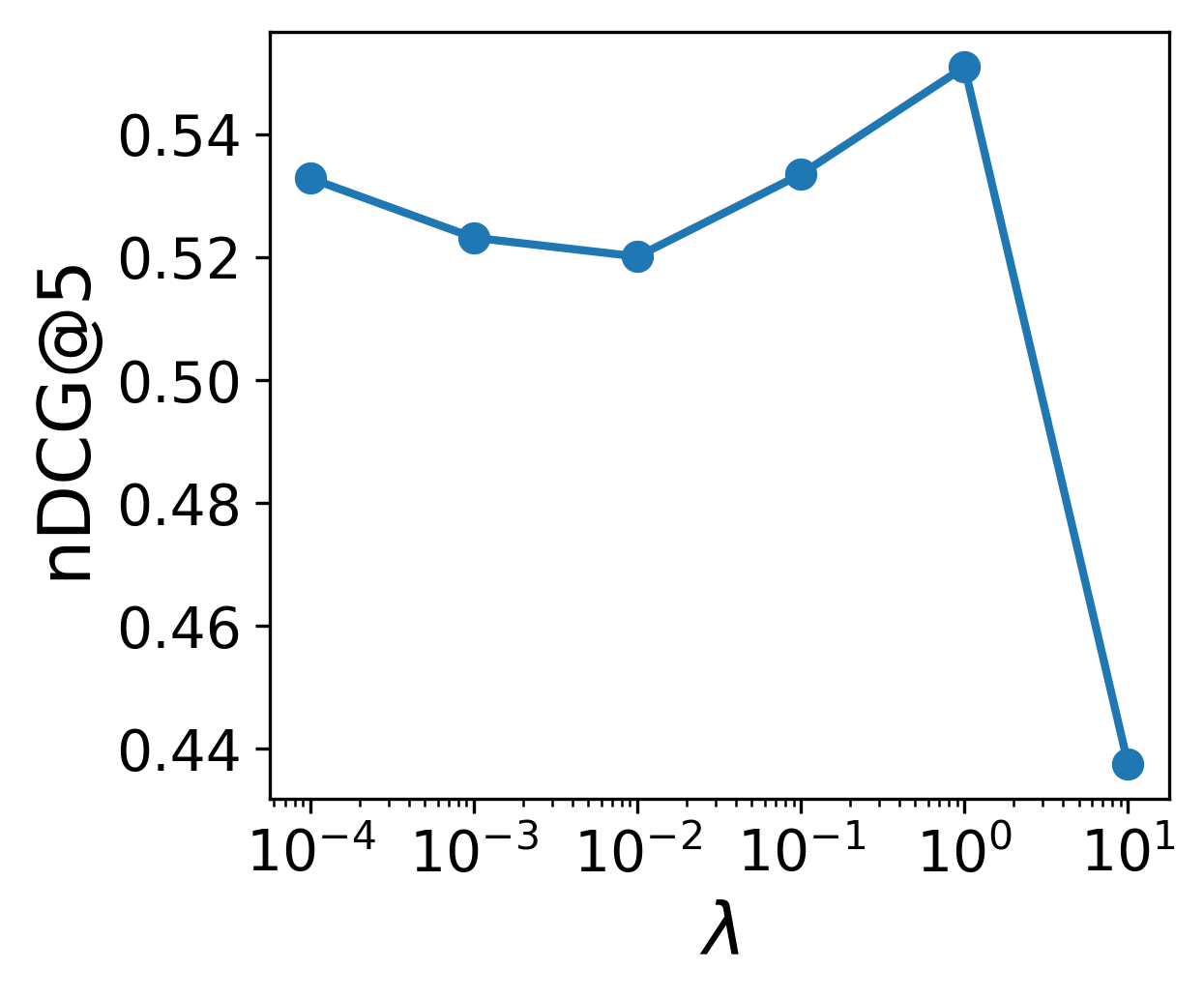}
        \caption{Effect of $\lambda$}
        \label{fig-lambda}
    \end{subfigure}
    \caption{Effect of different model parameters on the performance of CR-MLS}\label{fig-param}
\end{figure*}

\partitle{Impact of similarity scores.}
Figure \ref{fig-weights} shows the performance of CR-MLS when varying $\alpha_1$, $\alpha_2$, and $\alpha_3$, keeping in each run the other two parameters fixed. The best performance is achieved with geographical similarity weight at $\alpha_1 = 0.5$, review-based similarity weight at $\alpha_2 = 0.2$, and category-based similarity weight at $\alpha_3=0.3$. From Figure~\ref{fig-weights}, we can also see how much each of the similarity measures contribute to the overall performance. To do so, we take the performance of CR-MLS when the value of each $\alpha$ equals zero. This values indicates the performance of CR-MLS when the respective similarity measure is ablated from the model. More specifically, the performance of CR-MLS in terms of nDCG@5 ablating each of the similarity scores is as follows:
\begin{itemize}
    \item CR-MLS: nDCG@5 = 0.5509
    \item CR-MLS-NoGeographical: nDCG@5 = 0.5288
    \item CR-MLS-NoReview: nDCG@5 = 0.5375
    \item CR-MLS-NoCategory: nDCG@5 = 0.5306
\end{itemize}
We can see that the performance is dropped after removing each of the similarity scores, indicating that each of these similarity scores contribute to the overall performance of CR-MLS. Ablating the geographical similarity results in the highest drop compared to the other similarity measures. This shows that geographical similarity captures the similarity of two venues more effectively and reflects users' preference more accurately.

\partitle{Impact of number of latent factors.}
In \autoref{fig-d} we study the effect of latent factors $d$ on the performance of our model and report nDCG@5 while keeping other parameters fixed.
We observe that the optimal number of latent factors $d$ is 90, and for other values of latent factors, nDCG@5 drops. As we can see, the difference in the performance of the model is more when varying $d$ from 80 to 90. This indicates the importance of finding the optimal number of latent factors for training CR-MLS as it can have large impact on the model's performance since the number of latent factors are crucial while learning the latent associations between users and venues.

\partitle{Impact of regularization parameter.}
\autoref{fig-lambda} shows the effect of the regularizing control parameter ($\lambda$). We varied $\lambda$ while keeping other parameters fixed. We see that CR-MLS performs best with $\lambda=1$. The performance of CR-MLS drops using different other values of $\lambda$.

\section{Conclusions and Future Work}
\label{sec-conclusion}
In this paper, we presented a similarity-aware collaborative ranking framework for venue suggestion, called CR-MLS. 
The proposed CR-MLS is able to include an arbitrary number of cross-venue similarity measures in the model's objective function enabling the model to propagate venue affinities to the users and hence address the data sparsity problem.
To demonstrate the performance of CR-MLS, we also proposed three example cross-venue similarity measures focusing on different aspects. Geographical similarity incorporates the neighborhood influence of venues while category-based similarity takes into account venues that provide similar services. A review-based similarity score was also computed extracting an opinion- and context-based similarity of venues.  We compared the performance of CR-MLS with five collaborative and one content-based state-of-the-art approaches on a combined dataset of two publicly available TREC collections. The results indicated that our model can address the data sparsity problem, outperforming the state-of-the-art methods significantly. While we introduced three example similarity functions, it should be noted that CR-MLS is very flexible to incorporate various features.
As future work, we plan to study how other similarity features, such as personal tags, time-based similarity, and contextual information from multiple LBSNs can improve venue suggestion~\cite{DBLP:conf/ecir/Aliannejadi17}.  

\noindent \textbf{Acknowledgements.} 
This research was partially funded by the RelMobIR project of the \grantsponsor{}{Swiss National Science Foundation (SNSF)}{http://www.snf.ch/en/Pages/default.aspx}.

\bibliographystyle{ACM-Reference-Format}
\bibliography{main}

\end{document}